\documentclass[referee,useAMS,usenatbib,usegraphicx]{mn2e}
\usepackage{amssymb}
\usepackage{hyperref}

\newcommand{\cmt}[1]{}
\newcommand{\Rs}{R_{sub}}
\newcommand{\Rt}{R_{tot}}
\newcommand{\de}{\partial}
\newcommand{\pj}{p_{inj}}
\newcommand{\px}{p_{max}}
\newcommand{\degK}{~\rmn{ K}}
\newcommand{\ud}{\rmn{d}}
\newcommand{\xFP}{x_{A}}
\newcommand{\fo}{f_{sh}(p)}
\newcommand{\Pth}{P_g}
\newcommand{\apj}{ApJ}

\newcommand{\APh}{APh}
\def\ie{{\it i.e.,~}}
\def\kms{~{\rm km~s^{-1}}}
\def\cm3{~{\rm cm^{-3}}}

\title[Comparison of Different Methods for NLDSA]
{Comparison of Different Methods for Nonlinear Diffusive Shock Acceleration}

\author[Caprioli, Kang, Vladimirov and Jones]
{D.~Caprioli,$^{1}$\thanks{E-mail: caprioli@arcetri.astro.it},
H.~Kang,$^{2}$\thanks{E-mail: kang@uju.es.pusan.ac.kr},
A.~E.~Vladimirov,$^{3}$\thanks{E-mail: avladim@stanford.edu},
T.~W.~Jones,$^{4}$\thanks{E-mail: twj@msi.umn.edu},\\
$^{1}$INAF-Osservatorio Astrofisico di Arcetri, Largo E. Fermi, 5, 50125, Firenze, Italy\\
$^{2}$Department of Earth Sciences, Pusan National University, Pusan 609-735, Korea\\
$^{3}$Hansen Experimental Physics Laboratory, Stanford University, Stanford, CA 94305-4085, USA\\
$^{4}$Department of Astronomy, University of Minnesota, Minneapolis, MN 55455, USA\\}

\begin{document}
\date{\today}
\pagerange{\pageref{firstpage}--\pageref{lastpage}} \pubyear{2010}

\maketitle
\label{firstpage}

\begin{abstract}
We provide a both qualitative and quantitative comparison among different approaches aimed to solve the problem of non-linear diffusive acceleration of particles at shocks.
In particular, we show that state-of-the-art models (numerical, Monte Carlo and semi-analytical), even if based on different physical assumptions and implementations, for typical environmental parameters lead to very consistent results in terms of shock hydrodynamics, cosmic ray spectrum and also escaping flux spectrum and anisotropy. 
Strong points and limits of each approach are also discussed, as a function of the problem one wants to study.
\end{abstract}


\maketitle

\section{From DSA to NLDSA}
In the late 70's many authors \citep{krimskii,als77,b078,bell78} introduced the \textit{test-particle} theory of particle acceleration at strong collisionless shocks due to first order Fermi mechanism.
However, quantitative estimates soon pointed out that this mechanism, called Diffusive Shock Acceleration (DSA), may be so efficient that the back-reaction of the accelerated particles on the shock dynamics can not be neglected.

The obvious theoretical challenge is how to model effectively the full shock dynamics. 
Modeling shocks with efficient particle acceleration using plasma simulations (PIC and hybrid) is extremely computationally expensive for several reasons. First, the energies of charged particles participating in the process range from the low thermal energies of cold plasma to the ultra-relativistic energies of cosmic rays, and both extremes of particle spectra are dynamically important if acceleration is efficient \citep[see, e.g.,][for estimates of computational requirements]{veb08}. Second, simulations need to be done in three dimensions because of the possibility of non-physical suppression of important processes in one- and two-dimensional simulations \citep[see, e.g., ][]{jjb98}. Therefore, approximate methods must be used to model efficient particle accelerating shocks.

The first step towards a Non-Linear theory of DSA (NLDSA), in which particle acceleration and shock dynamics are calculated self-consistently, is represented by the two-fluid model proposed by \cite{dr_v80,dr_v81}. 
This approach is indeed very useful to study the dynamics of a shock modified by the presence of accelerated particles (treated as a diffusive, relativistic fluid) but by design it cannot provide any information about the spectrum of the cosmic rays (CRs) which are produced.

A \textit{kinetic} study of the NLDSA at astrophysical shocks also requires a statistical description of transport of superthermal particles. 
This relevant piece of Physics has been taken into account by adopting a number of rather different approaches; namely
\begin{itemize}
\item fully numerical simulations, in which a time-dependent diffusion-convection equation for the CR transport is solved in concert with coupled gas dynamics conservation laws, like in \cite{bell87, bv97, kj97, kjg02, kj05, kj06, kj07, za10};
\item stationary Monte Carlo numerical simulations of the full particle population, like in \cite{ee84,ee85,emp90,ebj95,ebj96,ed02,veb06};
\item semi-analytic solutions of the stationary \citep[or quasi-stationary, see][]{bac07} diffusion-convection equation coupled to the gas dynamic equations, like in \cite{malkov97,mdv00, blasi1, blasi2, ab05, ab06,cabv09,cab09}.
\end{itemize}

All these approaches make consistent predictions about the main consequences of the shock modification, i.e.\ the appearance of an upstream precursor created  by the CR pressure around the shock, which slows down the incoming fluid. 
As a consequence of this CR-induced velocity gradient, the fluid compression ratio is no longer 4 for any strong shock. Hence, the test-particle prediction that the spectrum of accelerated particles has to be $\propto p^{-4}$ is no longer valid.
Nevertheless, it is still true that the spectral slope mainly depends on the compression factor actually ``felt'' by CRs. Since the larger is its momentum, the farther from the shock a particle can diffuse, high (low) momentum particles probe a total compression ratio larger (smaller) than 4.
The resulting CR spectrum becomes rather concave, being harder (softer) than $p^{-4}$ at the higher (lower) energies.
For accurate reviews of CR-modified shocks the reader can refer to \cite{drury83,be87,je91,md01}.

In this paper we want, for the first time, to construct both  qualitative and quantitative comparisons among different approaches to NLDSA, in  order to highlight their analogies and differences.
In particular, we compare three state-of-the-art kinetic methods which are representative of the three categories listed above: the fully numerical calculation by \cite{kj07}, the Monte Carlo simulation by \cite{veb06} and the semi-analytical solution by \cite{cab09}.
The final goal of this work is to provide a useful tool illustrating which physical aspects of the problem can be taken into account within each approach, which assumptions are \textit{a posteriori} justified and consistent with any observational signature, and finally what is the computational effort required to tackle a given problem.

The present work is organized as follows: the next three sections are dedicated to the description of the three different techniques adopted to solve the problem of NLDSA. 
For each of them, the assumptions (e.g.\ time-dependence, isotropy of the CR distribution function) and the equations of the model, the injection of particles in the acceleration mechanism and the turbulent heating in the precursor are outlined.
In \S\ref{sec:benchmark} the results obtained within different approaches are compared, in terms of both hydrodynamics and CR distribution function.
We comment and conclude in \S\ref{sec:conclusion}.

\section{Numerical approach}\label{sec:KJ}
\subsection{Basic equations}
In their kinetic simulations of DSA, Kang and Jones (hereafter KJ) evolve the standard time-dependent gasdynamic conservation laws with the CR pressure terms in Eulerian form for one-dimensional plane-parallel geometry
\citep{kjg02,kj07},
\begin{eqnarray}
\frac{\partial \rho}{\partial t} &=& - \frac{\partial}{\partial x}(\rho u)\label{masscon}\\ 
\frac{\partial}{\partial t} (\rho u) &=& -\frac{\partial}{\partial x}
\left( \rho u^2 + P_g + P_c\right)\label{mocon}\\ 
\frac{\partial}{\partial t}\left( \frac{1}{2}\rho u^2 +
  \frac{P_g}{\gamma-1} \right) &=& 
-\frac{\partial}{\partial x}\left( \frac{1}{2}\rho u^3 +
  \frac{\gamma u P_g}{\gamma-1} \right) - u \frac{\partial P_{c}}{\partial
x}+\mathcal{W}(x,t)-\mathcal{L}(x,t)~\label{econ}.
\label{eq:conserva}
\end{eqnarray}
where $\rho=\rho(x,t)$ is the gas density, $u=u(x,t)$ is the fluid velocity, 
$P=P(x,t)$ is the gas ($P_{g}$) or CR ($P_{c}$) pressure (eq.~\ref{xicr})
and $\gamma= 5/3$ is the gas adiabatic index.
$\mathcal{L}(x,t)$ is the injection energy loss term, which accounts for the
energy carried away by the suprathermal particles injected into the CR
component at the subshock and is subtracted from the postshock gas
immediately behind the subshock.
The gas heating due to the Alfv\'en wave dissipation in the upstream region is modeled following \cite{mck-v82} and is given by the term 
\begin{equation}
	\mathcal{W}(x,t)= - v_A(x,t) \frac{\partial P_c}{ \partial x },
\label{wheat}
\end{equation}
where $v_A(x,t)= B_0/\sqrt{4\pi \rho(x,t)}$ is the local Alfv\'en speed
and $B_0$ is the magnetic field strength far upstream.
Here and in the following we will label with the subscript 0 quantities at upstream infinity,
and with 1 and 2 quantities immediately upstream and downstream of the subshock, respectively.
These equations can be used to describe parallel shocks, where the
large-scale magnetic field is aligned with the shock normal and
the pressure contribution from the turbulent magnetic fields
can be neglected.

The CR population is evolved by solving the diffusion-convection equation 
for the pitch-angle-averaged (isotropic in momentum space) distribution function, 
$f(x,p,t)$, in the form:
\begin{equation}\label{diffcon}
\frac{\de f}{\de t}+(u+u_{w})\frac{\de f}{\de x}=\frac{\de}{\de x}\left[D(x,p)\frac{\de f}{\de x}\right]+\frac{p}{3}\frac{\de f}{\de p}\frac{\de (u+u_{w})}{\de x}\,,
\end{equation}
where $D(x,p)$ is the spatial diffusion coefficient \citep[see e.g.][for a derivation]{ski75} and $p$ is the scalar, momentum magnitude.
This equation is then usefully rewritten and solved in terms of $g(x,p,t)=p^4f(x,p,t)$ and $y=\log(p)$ as explained in \cite{kjg02}, Eq.~7.
The CR population is isotropized with respect to the local Alfv\'enic
wave turbulence, which would in general move at a speed $u_w$ with respect
to the plasma.
Since the Alfv\'en waves upstream of the subshock are expected to be established
by the streaming instability, the wave speed is set there to be $u_w=-v_A$.
Downstream, it is likely that the Alfv\'enic turbulence is nearly
isotropic, hence $u_w=0$ there.

KJ typically initialize their simulations by creating 
pure hydrodynamical shocks and then evolving them along with CRs.
The shocks described here were initialized from Rankine-Hugoniot (R-H)
relations, although alternatives, such as reflecting boundaries
can be employed. It is important to note that the initial shock
speed established by the R-H conditions (and thus, also the Mach number)
will change as the shock evolves. 
That results from increased compression through the shock. This
feature, which is not a factor in the two steady state shock models,
complicates somewhat our comparison methodology.

Previous studies showed that CR modified shocks evolve to a dynamical self-similar state,
with constant total compression and downstream pressure balance; however, the CR
population continues to stretch to ever higher momenta unless high energy
CRs escape either through an upper momentum boundary, or a spatial ``free escape''
boundary \citep{krj09}. The former is simply implemented by setting $f(x,p) = 0$ for $p > p_{max}$,
while the latter (employed in this study) is effected by setting $f(x,p) = 0$
for $x > x_0$. Then the modified shocks and their CR population will
evolve until CR escape through either of these boundaries balances
DSA within the shock structure. From that time forward the shock and its
CR population become steady and the spectrum of escaping particles can be worked out numerically as
\begin{equation}\label{eq:escKJ}
\phi_{esc}(p)=-D(p)\left[\frac{\partial f}{\partial x}\right]_{x_0}\,.
\end{equation}

\subsection{Method of solution}
Equations \ref{masscon}, \ref{mocon}, \ref{econ}, and \ref{diffcon} are simultaneously 
integrated by finite difference methods in the CRASH (Cosmic-Ray Acceleration SHock) code. 
The detailed description of the CRASH code can be found in \citet{kjg02}. 
Three features of CRASH are important to our discussion below.
First, CRASH applies an adaptive mesh refinement technique around the subshock.
So the precursor structure is adequately resolved to couple
the gas to the CRs of low momenta, whose diffusion lengths can be
at least several orders of magnitude smaller than the precursor width. 
Second, CRASH uses a subgrid shock
tracking; that is, the subshock is a true discontinuity, whose position is followed accurately within a single
cell on the finest mesh refinement level. Consequently, the effective
numerical subshock thickness needed to compute the spatial derivatives
in Eq.~\ref{diffcon} is always less than the single cell size of the
finest grid. 
Third, the exact subshock speed is calculated at each time step to adjust the rest frame of the
simulation, so that the subshock is kept inside the same grid cell throughout. 
These three features allow one to obtain good numerical
convergence in the solutions with a minimum of computational effort.
A typical shock simulation with CRASH evolved to steady state with CRs
accelerated to $\sim 10^3$GeV/c requires roughly 300 CPU hours on
current microprocessors. CRASH does not currently have a parallel implementation.
The evolving shock simulation time can be shortened by as much as an order of magnitude
using the finite volume, CGMV algorithm for CR transport discussed in \cite{jk05} instead 
of traditional finite difference methods. However, thermal leakage
in the sophisticated form discussed in the following subsection has not been
implemented with the CGMV algorithm. So, since one of our objectives here is
comparisons of injection methods, we do not employ CGMV.

\subsection{Thermal leakage injection}\label{KJ:inj}
Since the velocity distribution of suprathermal particles in the Maxwellian tail is not 
isotropic in the shock frame, the diffusion-convection equation cannot directly 
follow the injection from the non-diffusive thermal pool into the diffusive CR population. 
In the CRASH code, thermal leakage injection 
is emulated numerically by adopting a \textit{transparency function}, 
$\tau_{\rm esc}(\epsilon_B, \upsilon)$,
which expresses the probability for downstream thermal particles at given random velocity, 
$\upsilon$, to swim successfully upstream across the subshock through the
postshock cyclotron-instability-generated MHD waves, whose amplitude is parameterized by $\epsilon_B$
\citep[see][Eq.~8]{kjg02}.
The condition of finite probability for suprathermal downstream particles to 
cross the subshock (i.e.\ $\tau_{\rm esc}>0$ for $p$ larger than an injection 
momentum $p_{\rm inj}\approx (1+ 1.07/\epsilon_B)m_p u_2$
effectively selects the lowest momentum of the particles entering the CR population.
The model transparency function increases smoothly from zero to unity over $p_{\rm inj}<p<
\sim 5 p_{\rm inj}$. 
Consequently, the transition from the Maxwellian to the quasi-power-law distribution develops smoothly in momentum space.
The parameter, $\epsilon_B = B_0/B_{\perp}$, is the ratio of the magnitude of the large-scale magnetic field
aligned with the shock normal, $B_0$, to the amplitude of the postshock 
wave field that interacts with low energy particles, $B_{\perp}$.
This is an adjustable parameter, although, as discussed in \cite{malkov97},
it is confined theoretically and experimentally to values, $\epsilon_B \sim 0.3$.

\section{Monte Carlo approach}\label{sec:EV}
\subsection{Model assumptions}
The method of solving the NLDSA problem based on the Monte Carlo particle transport, developed by Ellison and co-workers
and later adapted for modeling magnetic field amplification in shocks by Vladimirov (hereafter EV), makes the following physical assumptions:
\begin{itemize}
\item all quantities depend only on one spatial coordinate, but the flow velocity and/or the uniform magnetic field may have a component along the perpendicular direction;
\item a steady state is realized;
\item mass, momentum and energy conservation laws determine the nonlinear shock structure;
\item the transport of all particles may be described statistically.
  In the simplest case, it is assumed that the momentum vector of every particle
  gets randomly perturbed in the course of propagation in such a way
  that particle distribution is isotropized in the course of a momentum-dependent collision time, which
  is a parameter of the model. If scattering rate is constant in space and time, then on large
  enough scales such particle transport models diffusion. Gyration around a uniform magnetic 
  field may be included along with the stochastic component of the motion;
\item particles of all energies are treated on an equal basis, i.e.\ there
  is no need for a strict distinction between thermal and superthermal particles.
  However, such a distinction may be artificially imposed (see \S\ref{EV:th});
\item collisions that facilitate particle diffusion are elastic in the reference frame moving 
  at a speed $u_w$ with respect the thermal plasma, where $u_w$ is chosen the same way as in the other
  models.
\end{itemize}

The problem of NLDSA is thus stated as that of finding a steady state
particle distribution function in a system
where a supersonic motion exists at an infinite boundary, and the flow
is subsonic at the opposite boundary. 
The distribution function $F(x, {\bf p})$ in this case retains the information about
the direction of the momentum vector ${\bf p}$ and spans the range of momenta 
from thermal to $p_\mathrm{max}$.
Here and in the following we indicate with $F(x, {\bf p})$ the whole distribution function (thermal + cosmic rays)
and with $f(x, {\bf p})$ the cosmic ray distribution only.

\subsection{Method of solution}\label{EV:model}
The method by which this problem is solved involves an iterative
application of the following two steps:

1) a trial flow speed $u(x)$ is chosen based on the conservation laws and on the best-yet estimate 
of $F(x, {\bf p})$, and 

2) the updated particle distribution $F(x, {\bf p})$ is calculated with
the Monte Carlo technique based on the trial flow speed $u(x)$.\\
These steps are repeated until mass, momentum and energy conservation laws hold
for the particle distribution $F(x, {\bf p})$ calculated with the chosen $u(x)$.

In practice, the first step is realized by adjusting
the previous guess for $u(x)$ using the conservation laws. 
The conservation of momentum is used to choose $u(x)$ in
the supersonic region (i.e., upstream), and the conservation of 
momentum and energy 
is used to choose the flow speed in the subsonic region (i.e.\ downstream), taken as uniform. 
The equations expressing the conservation of mass, momentum and energy fluxes can be written
in terms of $F(x,{\bf p})$ as
\begin{eqnarray}
\int m_p v_x F(x, {\bf p})~d^3p &=& \rho_0 u_0 \\
\int p_x v_x F(x,{\bf p})\ud^3p &=& \rho_0 u_0^2 + P_{g,0}~,\label{pflux}\\
\int K({\bf p}) v_x F(x,{\bf p})\ud^3p &=& \frac12 \rho_0 u_0^3 + \frac{\gamma}{\gamma-1}u_0 P_{g,0} - Q_\mathrm{esc}~,\label{eflux}
\end{eqnarray}
where the subscript $x$ labels the component along the axis $x$, $K({\bf p})$ is 
the kinetic energy of a particle with momentum ${\bf p}$ and $Q_\mathrm{esc}$ 
is the energy flux carried away by particles escaping the system (see below for details).
This way of dealing with a \textit{global} distribution function is in contrast to 
writing the left hand side as the sum of gas and CR pressures, which would require 
$F(x,{\bf p})$ to be isotropic in $p$-space. 

The second step is realized by introducing a population of thermal particles
either far upstream, or at a location $x_\mathrm{A}$ (defined below), and
propagating them in time steps small compared to the collision time, 
until they escape. A particle is considered to have escaped if
a) it is found many diffusion lengths downstream, or b) it crosses
the upstream free escape boundary. 
As the particles are propagated, their momentum are randomly perturbed as 
described in the next paragraph. 
When particles cross selected coordinates $x_j$, their momenta and
statistical weights are summed to calculate the particle distribution function
in a momentum bin:
\begin{equation}
F(x_j, p_k, \mu_l) = 
 \sum_{p \in \Delta p_k}\sum_{\mu \in \Delta \mu_l} 
       \frac{\sigma_i}{2 \pi p_k^2 \Delta p_k \Delta \mu_{l}}.
\end{equation}
Here, the summation takes place over the events of a particle crossing
the coordinate $x_j$: $\Delta p_k$ is a bin in momentum space, 
$\Delta \mu_l$ is a bin in angles ($\mu = p_x / p$) and $\sigma_i$
is the properly normalized statistical weight of the particle.
In a similar fashion, the code calculates the moments of the distribution function,
including the momentum and energy fluxes defined in Eq.~\ref{pflux} and \ref{eflux}. 

When particles are propagated, they move in straight lines for a time 
$\Delta t$ equal to a fraction of the collision time $t_c$. 
After that, the momentum in the local plasma frame is rotated by a random angle chosen
so that the averaged particle motion represents diffusion with the
appropriate diffusion coefficient. 
The local plasma frame for CR particles is assumed to be moving at a speed $u(x)-v_A(x)$ with respect 
to the subshock, where $v_A(x)=B_0/\sqrt{4\pi \rho(x)}$ is the local
Alfv\'{e}n speed. 
For thermal particles, the frame in which momentum rotation is performed is moving at a speed $u(x)$ with
respect to the subshock.
The details of this procedure are described in \cite{andrey}.
In regions where the particle distribution is isotropic (e.g., far
downstream), the code may set $\Delta t = t_c$, which simulates
a random walk process and corresponds to the diffusion approximation.

A gradient of the local plasma speed in the above procedure leads to 
the first order Fermi acceleration of particles. This applies to the thermal
particles as well as to the superthermal ones. Injection of thermal particles
into the acceleration process occurs in this model only due to
their Fermi-I acceleration at the subshock. 

Elastic scatterings of thermal particles in the shock precursor
will, in the presence of a small flow speed gradient, lead to adiabatic
heating: $\Pth \propto \rho^{\gamma}$ (this heating is, in a sense,
due to the same Fermi-I acceleration process, but on a smooth
variation of the flow speed instead of a discontinuity). 
If additional heating operates in the precursor due to some wave
dissipation processes, the treatment of thermal particles
in the precursor is modified as described in \S\ref{EV:th}.

In the Monte Carlo particle transport routine, the escape of energetic
particles upstream is modeled by removing from the simulation the particles that
cross the coordinate of the free escape boundary, $x_0$, in the upstream direction. 
The value of $x_0$ is a free parameter of the model.
The energy flux carried away by the escaping particles is reflected in
the value of the energy flux downstream of $x_0$.
The amount of escaping energy flux, $Q_{esc}$, is calculated in the EV method as the
difference between the energy flux upstream of $x_0$ and the
energy flux downstream of $x_0$ (see equation (\ref{eflux})), while
the spectral density of the flux of escaping particles can be calculated in the simulation as
\begin{equation}\label{eq:escEV}
\phi_{esc}(p)=\int_{-1}^{+1}\ud \mu ~c \mu F(x_0, p, \mu).
\end{equation}

As a final remark, the computational time required by this Monte Carlo approach
is on the order of tens to hundreds of processor hours, and parallel processing
is implemented in the latest version.

\subsection{Precursor heating}\label{EV:th}
In order to include precursor heating, particles are introduced
at an upstream location $\xFP$, which is close to the subshock,
but not too close, in order that the thermal particle distribution is still
isotropic at $\xFP$.
Upstream of $\xFP$, instead of calculating $F(x,{\bf p})$ and
its moments like $\Pth(x)$ from particle propagation, the model
finds $\Pth(x)$ by solving the equation
\begin{equation}\label{turbheat}
u(x) \frac{\rho^{\gamma}(x)}{\gamma-1}\frac{d}{dx}\left[ \frac{\Pth(x)}{\rho^{\gamma}(x)} \right] = \mathcal{W}(x)=-v_{A}(x)\frac{dP_{c}}{dx},
\end{equation}
where $\mathcal{W}(x)$ is the same heating rate as in Eq.~\ref{wheat}. 
Then particles are introduced at $\xFP$ with the temperature corresponding to $\Pth(\xFP)$
and allowed to propagate, get injected and accelerated. 

This procedure requires a differentiation between thermal and
superthermal particles in order to calculate the total momentum
and energy fluxes upstream (Eqs.~\ref{pflux} and \ref{eflux}). 
Having placed the shock at $x=0$ (with the upstream at $x>0$),
the definition adopted in the model is that a particle is thermal if it has never crossed
the coordinate $x=0$ going against the flow (\ie it never
got injected), and a CR particle otherwise. 
Therefore, for $x>\xFP$ Eqs.~\ref{pflux} and \ref{eflux} are rewritten as
\begin{eqnarray}
\rho(x) u^2(x) + \Pth(x) + 
        \int p_x v_x f(x,{\bf p})\ud^3p &=& \rho_0 u_0^2 + P_{g,0},\\
\frac12 \rho(x) u^3(x) + \frac{\gamma u\Pth(x)}{\gamma-1} + 
        \int K v_x f(x,{\bf p})\ud^3p &=& \frac12 \rho_0 u_0^3 + \frac{\gamma u_{0}P_{g,0}}{\gamma-1} - Q_\mathrm{esc},
\end{eqnarray}
where $f(x,{\bf p})$ is the distribution function containing only cosmic
ray contribution in the sense of the above definition. At $x>\xFP$, where anisotropies
of thermal particle distribution are important, Eqs.~\ref{pflux} and \ref{eflux} are used. 

As one can see, between the location $\xFP$ and the subshock, 
thermal particles are only heated by compression, but if this 
compression is rapid, like at the
subshock, this compressive heating is non-adiabatic. In fact, it turns out
that the solutions provided by the model contain a region of finite thickness
$0 > x > x_\mathrm{crit}$, in which this non-adiabatic compressive heating
takes place: this region is interpreted as the subshock.

\section{Semi-analytic approach}\label{sec:CAB}
The semi-analytic formalism for the NLDSA problem developed by Caprioli, Amato and Blasi 
(hereafter CAB) couples the conservation of mass, momentum and energy flux with an analytical solution of the stationary diffusion-convection equation for the isotropic part of the CR distribution function \citep{cab09}.

More precisely, the transport equation is taken as in Eq.~\ref{diffcon}, with the velocity of the scattering centres $u_{w}$ neglected with respect to the fluid one, their ratio being typically of order of $v_{A}/u\ll 1$:
\begin{equation}\label{diffconvstat}
u\frac{\de f}{\de x}=\frac{\de}{\de x}\left[D(x,p)\frac{\de f}{\de x}\right]+\frac{p}{3}\frac{ du}{dx}\frac{\de f}{\de p}+\mathcal{Q}(x,p)\,.
\end{equation}
The injection term $\mathcal{Q}(x,p)$ is written following \cite{bgv05} as  
\begin{equation}\label{eq:eta}
\mathcal{Q}(x,p)=\frac{\eta n_{0}u_{0}}{4\pi \pj^{2}}\delta(p-\pj)\delta(x)\,,
\end{equation}
where $\eta$ is the fraction of particles crossing the shock and injected in the acceleration process and $\pj$ is the injection momentum (see \S\ref{CAB:inj} for further details).

Eq.~\ref{diffconvstat} is solved imposing the upstream boundary condition $f(x_{0},p)=0$ as in KJ model, which mimics the presence of a free-escape boundary at a distance $x_{0}$ upstream of the shock.
A net flux of particles $\phi_{esc}(p)$ escaping the system from this boundary is expected to grant the achievement of a stationary configuration, as discussed in \cite{escape}.

An excellent approximate solution of Eq.~\ref{diffconvstat} is found to be:
\begin{eqnarray}
f(x,p)&=&\fo\exp\left[-\int_{x}^{0} \ud x'\frac{u(x')}{D(x',p)}\right]	\left[ 1-\frac{\Lambda(x,p)}{\Lambda_{0}(p)}\right];\label{eq:app-f}\\
\phi_{esc}(p)&=&-D(p)\left[\frac{\partial f}{\partial x}\right]_{x_0}=- \frac{u_{0}\fo}{\Lambda_{0}(p)}~,\label{eq:escCAB}
\end{eqnarray}
where
\begin{equation}
\Lambda (x,p)=\int_{x}^{0} \ud x' \frac{u_{0}}{D(x',p)}\exp\left[\int_{x'}^{0}\ud x''\frac{u(x'')}{D(x'',p)}\right];\qquad  \Lambda_{0}(p)=\Lambda(x_{0},p)
\end{equation}
embeds all the information on the escape flux and the CR distribution function at the shock position $f_{sh}(p)$ can be implicitly written as
\begin{eqnarray}
\fo &=&\frac{\eta n_{0}}{4\pi \pj^{3}}\frac{3\Rt}{\Rt U_{p}(p)-1}
	\exp\left\{-\int_{\pj}^{p}\frac{\ud p'}{p'}\frac{3\Rt}{\Rt U_{p}(p')-1}
	\left[U_{p}(p')-\frac{\phi_{esc}}{u_{0}f_{sh}(p')}\right]\right\},\label{eq:fsh}\\
U_{p}(p)&=&\frac{\Rs}{\Rt}-\frac{1}{\fo}\int_{x_{0}}^{0}\ud x f(x,p)\frac{dU}{d x}.\label{eq:Up}
\end{eqnarray}
Here we introduced the normalized fluid velocity $U=u/u_{0}$ and the compression ratios at the subshock $\Rs=u_{1}/u_{2}$ and between $x_{0}$ and downstream $\Rt=u_{0}/u_{2}$.

The solution of the problem is found using a recursive method, starting from a guess for $U_{1}=\Rs/\Rt$ (the compression ratios are linked together by Eq.~\ref{rsrtTH}), along with a guess for the CR distribution function at the shock and the escape flux \citep[the test-particle solution of][]{escape}.
\begin{itemize}
\item At each step of iteration the pressure in CRs is calculated from the distribution function as:
\begin{equation}\label{xicr}
	P_{c}(x)=\frac{4\pi}{3}~\int_{\pj}^{+\infty}\ud p ~p^3 ~v(p) ~f(x,p),
\end{equation}
and it is normalized to the valued of $P_{c,1}$ deduced from the momentum conservation equation integrated between positions immediately upstream of the subshock and $x_{0}$; namely
\begin{equation}
\frac{P_{c,1}}{\rho_{0}u_{0}^{2}}=1+\frac{1}{\gamma M_{0}^{2}}-U_{1}-P_{g,1}(U_{1}),
\end{equation}
where $P_{g,1}(U_{1})$ is given by Eq.~\ref{eq:Pgas}.
\item With this normalized CR pressure, it is possible to calculate at first $U(x)$ from momentum conservation and eventually a new $f(x,p)$ via Eqs.~\ref{eq:app-f}--\ref{eq:Up}.
\end{itemize}
The steps above are iterated, leaving $U_{1}$ unchanged until convergence is reached, i.e.\ until the  factor normalizing $P_{c,1}$ does not change between one step and the following: in general this factor is different from 1, which means that the chosen $U_{1}$ does not provide a physical solution of the problem.
The whole process is thus restarted with a new choice of $U_{1}$ until the iteration on $f(x,p)$ leads to a  distribution function which does not need to be re-normalized in order to satisfy momentum conservation: the $f(x,p)$ and the $U(x)$ found in this way represent the solution of the NLDSA problem.

Once the solution has been found, the equation for the conservation of the energy is satisfied as well, and the energy carried away by escaping particles, normalized to $\rho_{0}u_{0}^{2}/2$, can be written as
\begin{equation}\label{eq:Fesc-CAB1}
F_{esc}=\int_{\pj}^{+\infty}\ud p~\frac{8\pi p^2 K(p) \phi_{esc}(p)}{\rho_{0}u_{0}^{2}}\,.
\end{equation}
or also as the difference between the gas energy flux at $x_{0}$ and the whole energy flux downstream, namely:
\begin{equation}\label{eq:Fesc-CAB2}
F_{esc}=1-\frac{1}{\Rt^{2}}+\frac{2}{M_{0}^{2}(\gamma-1)}-\frac{2}{\Rt}\frac{\gamma}{\gamma-1}\frac{P_{g,2}}{\rho_{0}u_{0}^{2}}-\frac{2}{\Rt}\frac{\gamma_{c}}{\gamma_{c}-1}\frac{P_{c,2}}{\rho_{0}u_{0}^{2}}\,,
\end{equation}
where $\gamma_{c}\simeq 4/3$ is the adiabatic index of the relativistic gas of cosmic rays.

The time required to run this iterative scheme is less than one minute on a current generation CPU.

\subsection{Injection mechanism}\label{CAB:inj}
Also in this approach the injection of particles into the acceleration mechanism is taken as occurring immediately downstream of the shock due to thermal leakage, as in KJ.
However, the implementation by \cite{bgv05} adopted here is simplified with respect to the KJ model, because it does not include a smooth transparency function.

The injection momentum $\pj$ is in fact chosen as a multiple of the thermal momentum of  downstream particles, namely: $\pj=(\xi_{inj}-u_{2}/c)p_{th,2}$, with $p_{th}=\sqrt{2m_{p}k_{B}T_{2}}$. 
The shift $u_{2}/c$ comes from the assumption that thermal particles downstream have a Maxwellian spectrum in the fluid reference frame, while $\pj$ and the equations above are written in the shock frame.  
Under this hypothesis the fraction $\eta$ (Eq.~\ref{eq:eta}) is uniquely determined by the choice of $\xi_{inj}$, namely:
\begin{equation}
\eta=\frac{4}{3\sqrt{\pi}}\left(\Rs-1\right)\xi^{3} e^{-\xi^{2}}.
\end{equation}

\subsection{Turbulent heating}
As in the previous approaches, precursor heating due to the dissipation of Alfv\'enic turbulence can be taken into account as described in \cite{ab06}, following the implementation proposed by \cite{be99}.
In particular, the entropy conservation for the gas (Eq.~\ref{turbheat}) determines the gas pressure in the presence of a non-adiabatic heating as 
\begin{equation}\label{eq:Pgas}
	P_g(x) \simeq P_{g,ad}(x)\left[1+H(x)\right]\,;\qquad
	H(x)=\gamma(\gamma-1)\frac{M_0^2}{M_{A,0}}
	\left[\frac{1-U^{\gamma+1/2}(x)}{\gamma+1/2}\right],
\end{equation}
where $P_{g,ad}(x)/(\rho_{0}u_{0}^{2})=U^{-\gamma}(x)/(\gamma M_{0}^{2})$ is the plasma pressure as calculated taking into account only adiabatic compression in the precursor, i.e.\ $P_{g,ad}(x)/\rho^{\gamma}(x)=$const.
This expression serves as an equation of state for the gas in the
presence of effective Alfv\'en heating and leads to the following relation between the compression ratios $\Rs$ and $\Rt$ \citep{blasi1}:
\begin{equation}\label{rsrtTH}
\Rt^{\gamma+1}=\frac{M_0^2\Rs^\gamma}{2}\left[\frac{\gamma+1-\Rs(\gamma-1)}  
{(1+H_{1})}\right]\,.
\end{equation}
and in turn
\begin{equation}
\Rt=\frac{\gamma+1}{2U_{1}^{\gamma}(1+H_{1})/M_{0}^{2}+U_{1}(\gamma-1)};
\qquad \Rs=U_{1}\Rt
\end{equation}
where $H_{1}$ is given by eq.~\ref{eq:Pgas} evaluated upstream of the subshock.

For completeness, the downstream temperature is thus calculated to be
\begin{equation}
T_{2}=T_{0}\left(\frac{\Rt}{\Rs}\right)^{\gamma-1}
	\frac{\gamma+1-(\gamma-1)\Rs^{-1}}{\gamma+1-(\gamma-1)\Rs}~(1+H_{1})\,,
\end{equation}

\section{A benchmark case}\label{sec:benchmark}
In this section the three approaches described above are applied to the same physical problem, in order to emphasize analogies and differences among themselves.
The benchmark case studied here is a plane, parallel shock with velocity $u_{0}=5000\kms$ propagating in a homogeneous background with magnetic field $B_{0}=3\mu$G, particle density $\rho_{0}/m_{p}=0.003$ cm$^{-3}$ and temperature $T_{0}=2.02\times 10^{6}\degK$, corresponding to a sonic Mach number $M_{0}=30$.
The corresponding Alfv\'en speed is, $v_A \approx 120\kms$, with an Alfv\'enic Mach number, $M_A \approx 42$.
The upstream free escape boundary is located at $x_{0}=3.13\times10^{16}$cm and the diffusion is taken as Bohm-like diffusion in the background magnetic field $B_{0}$.
In approaches where a diffusion coefficient is needed, this is equivalent to
\begin{equation}
D(p) = D_{*}{p \over {m_pc}}~, 
\label{eq:D}
\end{equation}
with $D_* = 1.043\times 10^{22}{\rm cm^2 s^{-1}}(B_0/3 \mu{\rm G})^{-1}$.
In order to give an idea of the length-scale imposed here, the choice of $x_{0}$ corresponds to the diffusion length $\lambda=D(p)/u_{0}$ of a particle with momentum $p\simeq 10^{3}$GeV/c, which approximates the upper cutoff momentum,
$p_{max}$, obtained in all the solutions.

This set of environmental parameters is well defined for the stationary approaches of EV and CAB, but is slightly trickier for the time-dependent case of KJ. 
As noted previously, their simulations start with a purely gasdynamic 
shock established by R-H conditions to give a shock speed, $u_{0}=5600 \kms$ at rest 
at $x = 0$. 
There are no pre-existing CRs, \ie $P_c(x)=0$ at $t=0$.
As suprathermal particles are injected at the subshock and accelerated,
the CR pressure increases, while total compression increases and the shock is slowed down. 
When the shock has reached the self-consistent dynamical equilibrium state,
the speed becomes $u_{0}\approx 5000\kms$, which equals to the shock speed
of steady state solutions obtained by the other two methods. 
We emphasize again that a gasdynamic shock evolved into a CR modified
shock can reach a true steady state only when CRs escape the shock structure
and carry away energy at a rate that balances the rate at which CRs gain
energy through DSA. In these simulations that escape comes through imposition
of a free escape boundary at $x_0$.

In EV approach, there is no distinction between thermal and non-thermal particles, 
hence particle injection is intrinsically defined by the choice of the scattering properties (as in \S\ref{EV:model}),
and so it is not controlled with a free parameter. 
In order to match this intrinsic recipe for the injection embedded in the Monte Carlo approach,
we chose values of $\epsilon_B$ and $\xi$, which lead to similar total compression ratios
as EV's, $R_{tot}=\rho_2/\rho_0$.
Quantitatively speaking, KJ chose $\epsilon_B=0.28$ (see \S\ref{KJ:inj}), while CAB fixed $\xi=3.1$ (\S\ref{CAB:inj}), corresponding to the injection of a fraction $\eta\simeq 3\times10^{-3}$ of the flux of particles crossing the subshock.
It turns out these choices push the CR acceleration close to its maximum efficiency,
because the non-linear feedback and shock modification saturate with respect to increased 
injection.
It has been shown previously that, for moderately strong shocks ($M\ga 30$), the CR acceleration 
efficiency and the shock modification become insensitive to the values of $\epsilon_B$ and 
$\xi$ for the injection fraction above a critical value, $\eta \ga 10^{4}$
\citep[e.g.,][]{kj07}. 

\subsection{Particle spectrum}

\begin{figure}
   \centering
   \includegraphics[width=0.90\textwidth]{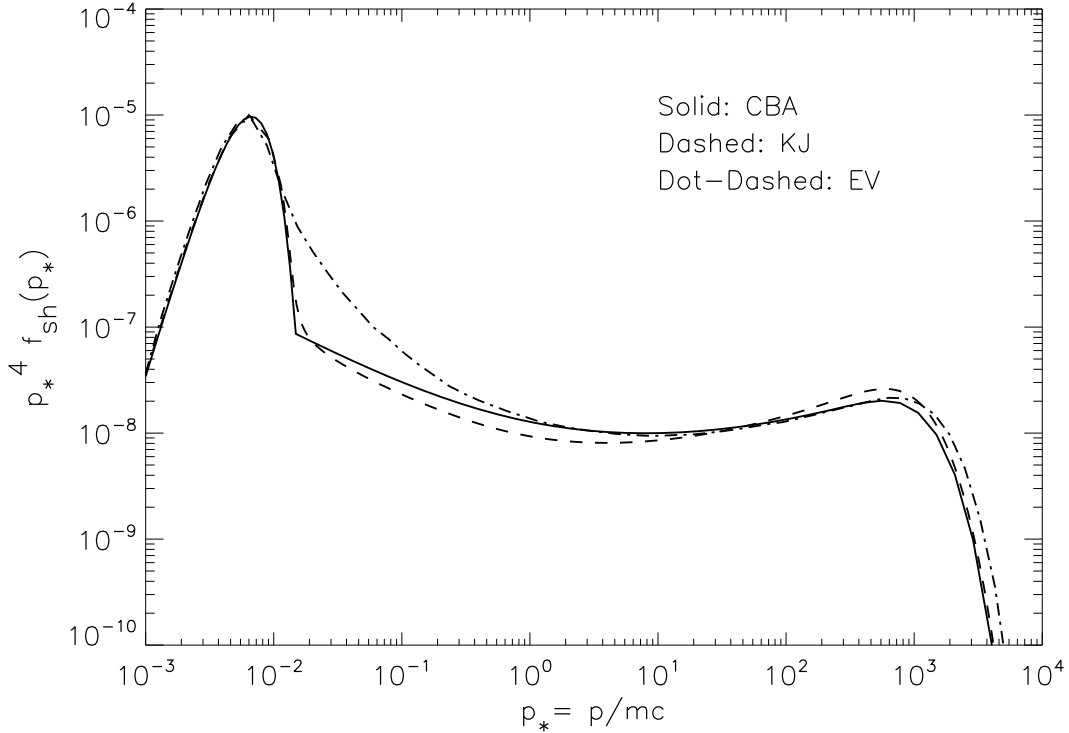}
  \caption{Particle distribution $f_{sh}(p/mc)$ at the shock position, in the shock frame, multiplied by $(p/mc)^{4}$,  calculated within different approaches as in the legend.}   
   \label{fig:spectra}
\end{figure}
In Fig.~\ref{fig:spectra} the particle distribution at the shock position is shown as a function of the normalized momentum $p_{*}=p/mc$, as multiplied by $p_{*}^{4}$ in order to emphasize the main scaling. 
The agreement among the three solutions is indeed very satisfactory: the concave shape of the CR spectrum typical of CR modified shock models is recovered within all the approaches: 
in particular, $f(p)$ is steeper than $p^{-4}$ at low momenta and then
flattens towards higher momenta, until it turns down sharply above $p_{max}$.

The peaks of the distributions of the thermal particles calculated with the three different approaches match very well, and the thermal distributions in the two thermal leakage injection models are in excellent agreement. 
In the Monte Carlo approach the transition between thermal and non-thermal particles is much broader than in thermal leakage ones. Moreover, the  transparency function implemented in KJ thermal leakage modeling makes this transition smoother than in CAB one, which shows a sharp break between thermal and non-thermal populations.
Nevertheless, despite of these different ways of dealing with particle injection, above 0.1 GeV/c all spectra are very close to each other.

As estimated at the beginning of the section, the spectrum has to be cut-off around $\px\approx10^{3}$GeV/c as a consequence of the escape of the highest energy particles, which are not able to diffuse back to the shock.
The position and the shape of the momentum cut-off is consistent among the different models, with the EV solution extending about 10 per cent higher in energy with respect to the other cases (see also the left panel of Fig.~\ref{fig:escape}). 
This is due to a subtle difference in the definition of the transport properties of particles: the models of CBA and KJ use a diffusion coefficient to describe relativistic particle propagation, and the model of EV defines a small-angle scattering rate instead, which leads to some 20 per cent difference in the effective particle mean free path, which is a small factor compared to the uncertainty in the model of particle diffusion \citep[see e.g.~\S~3.1.2 in][]{andrey}.

\subsection{Escape flux: spectrum and anisotropy}\label{sec:escape}

\begin{figure}
   \centering
   \includegraphics[width=0.48\textwidth]{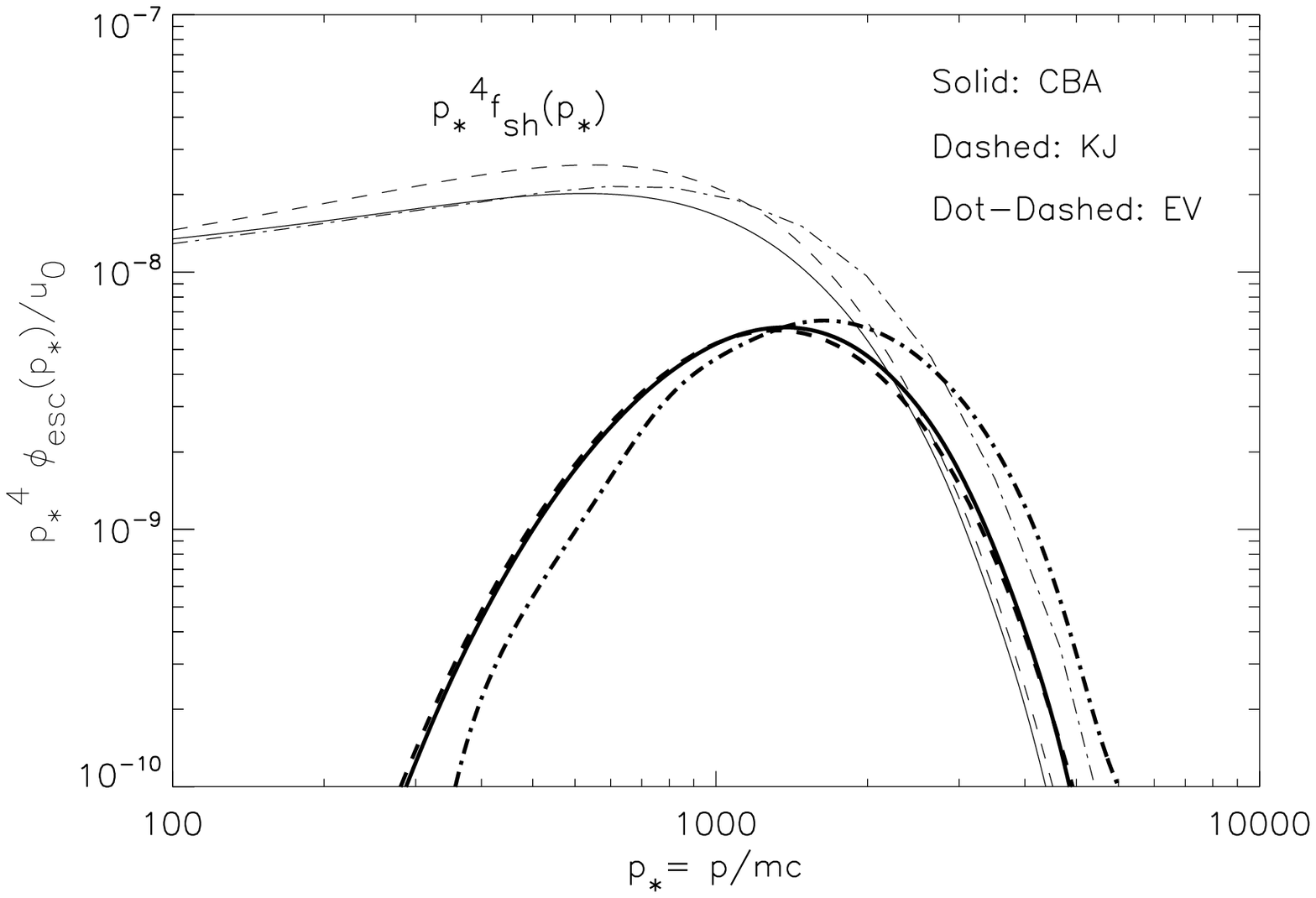}
   \includegraphics[width=0.48\textwidth]{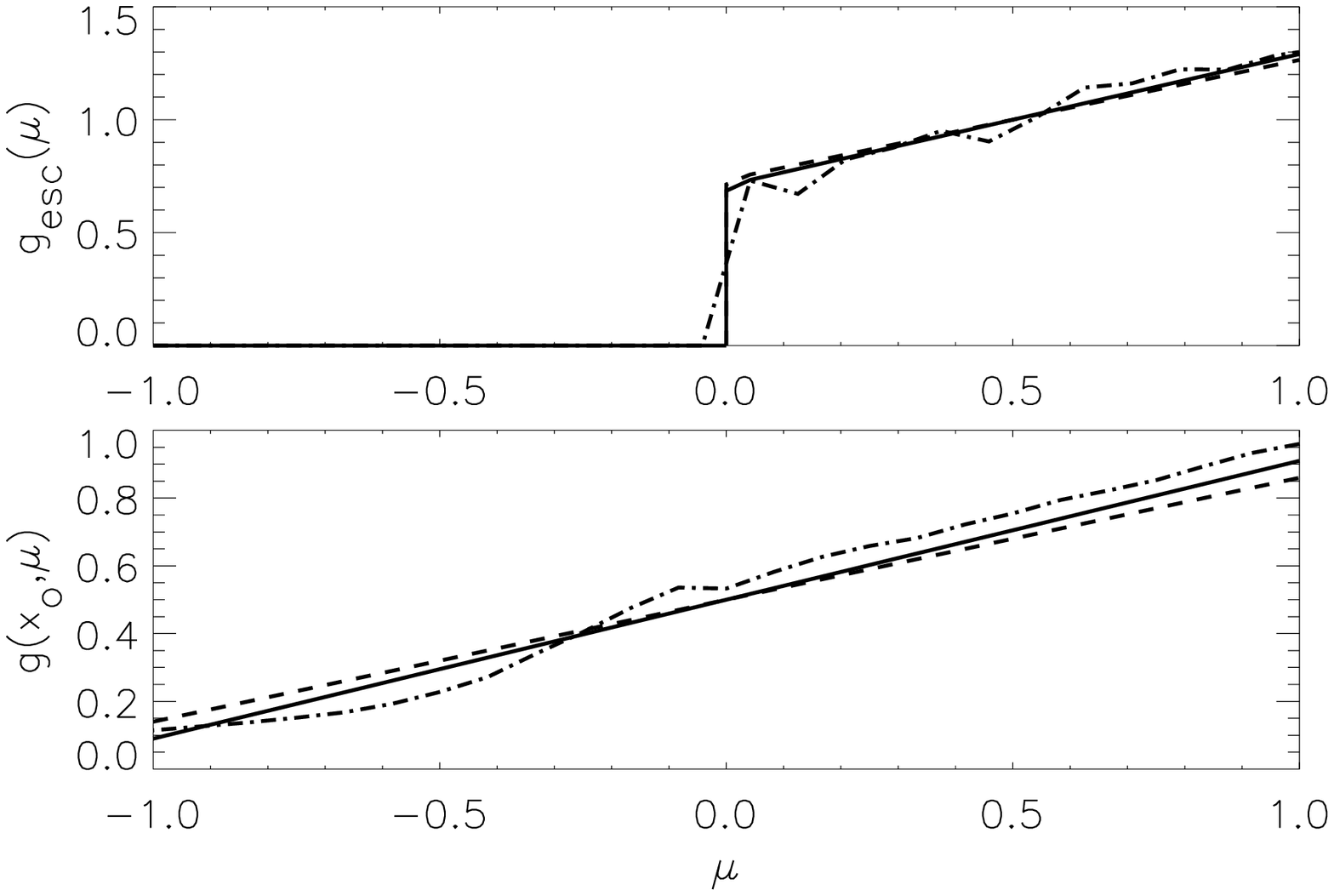}
   \caption{\textit{Left panel}: flux of particles escaping the system from the free-escaping boundary at $x_{0}$ ($\phi_{esc}$, heavy lines) and particle distribution at the shock ($f_{sh}$, thin lines). Different curves correspond to different models, as in the legend of Fig.~\ref{fig:spectra}.
   \textit{Right panel}: angular distribution, averaged over momenta, of escaping particles ($g_{esc}(\mu)$, top) and of the particles  immediately downstream of the free-escape boundary ($g(x_{0},\mu)$, bottom).}
   \label{fig:escape}
\end{figure}

The spectrum of particles escaping the system from the upstream boundary $\phi_{esc}(p)$ is shown in the left panel of Fig.~\ref{fig:escape} (heavy lines), as multiplied by $p_{*}^{4}/u_{0}$ and compared with the highest end of the spectrum at the shock location, $p_{*}^{4}f_{sh}(p)$ (thin lines).
The spectra obtained with different methods are in good agreement, with a slight shift towards higher energies of the peak of the spectral distribution found within the Monte Carlo approach as discussed above.

This fact is of great interest since particles escaping the system from the upstream are expected to be responsible for the highest energy region of the galactic diffuse cosmic ray spectrum \citep[see e.g.][]{cab10}. 
In particular, provided that escape occurs when the shock evolution is quasi-stationary (a reasonable assumption during the Sedov-Taylor stage of a SNRs, when the instantaneous maximum energy is thought to be limited by the accelerator size and not by the acceleration time), the rate and the spectrum of accelerated particles that are injected in the Galaxy as CRs is found to be the same in all the approaches.

This remarkable agreement between models that solve the particle transport via Monte Carlo techniques or via the diffusion equation is not trivial at all, since the anisotropy of the particle distribution function is treated in radically different ways.
In order to better understand this point we introduce the quantity 
\begin{equation}\label{eq:g}
g(x,\mu)=\frac{\int_{0}^{\infty}4\pi p^2 \ud p f(x, p, \mu)}{\int_{-1}^{+1}\ud\mu \int_{0}^{\infty}4\pi p^2 \ud p f(x, p, \mu)}\,,
\end{equation}
which measures the anisotropy of the CR distribution function, averaged over $p$, at the position $x$.
Here $\mu=p_{x}/p$, with $\mu=+1(-1)$ 
corresponding to particles moving against (in the direction of) the flow, 
and $g(x,\mu)$ is normalized such that $\int_{-1}^{+1}g(x,\mu)\ud \mu =1$.

While the Monte Carlo approach directly takes the anisotropy of the distribution function into account (see \S\ref{EV:model}), the other approaches solve the diffusion-convection equation for the isotropic part only. 
More precisely, in the diffusive approximation the distribution function is taken as the sum of an isotropic term $f^{(0)}(p)$ and a dipole-like anisotropic one $\mu f^{(1)}(p)$, in such a way that
\begin{equation}\label{eq:expansion}
 f(x,{\bf p})\simeq f^{0}(x,p)+\mu f^{1}(x,p)
\end{equation}
This is expected to be a very good approximation for non-relativistic shocks since the $n^{th}$ term in the anisotropy expansion is of order $(u/c)^{n}$ \citep[see e.g.][for a thorough derivation of the diffusion-convection equation]{ski75}. 
The introduction of the spatial diffusion coefficient in fact allows one to eliminate $f^{1}(x,p)$, and the diffusive flux reads 
\begin{equation}\label{eq:f1}
-D(p) \frac{\partial f^{0}(x,p)}{\partial x} = \int_{-1}^{+1}f({\bf p})c \mu \ud \mu = \frac23 c f^{1}(x, p).
\end{equation}

We are particularly interested in the anisotropy at the free-escape boundary, where it is expected to be the largest, but there is a caveat in carrying out this analysis.
Whichever is the value of the distribution function at $x_{0}$ as a function of $p$, the anisotropy immediately downstream of $x_{0}$, which we will refer to as $g(x_{0},\mu)$, and immediately upstream of $x_{0}$, indicated with $g_{esc}(p,\mu)$, are intrinsically different. This must be true, since upstream of $x_{0}$ particles must escape the system, and hence, $g_{esc}(p,\mu)$ has to vanish for $\mu<u_{0}/c\ll 1$. 
Downstream of $x_{0}$, instead, particles can diffuse back to the shock and thus Eq.~\ref{eq:expansion} safely holds also for $\mu<0$. 

Keeping this difference in mind, we can write the anisotropy of the distribution function at the position $x_{0}$ also for diffusive approaches using Eqs.~\ref{eq:g}-\ref{eq:f1}, obtaining
\begin{equation}\label{gmu}
 g(x_{0},\mu)=\frac{1}{2}\left(1+A\mu\right)\,; \qquad
 A=\frac{\int_{0}^{\infty}\ud p~  p^{2}f^{1}(x_{0},p)}{\int_{0}^{\infty}\ud p ~ p^{2}f^{0}(x_{0},p)}
  =\frac{3}{2c}\frac{\int_{0}^{\infty}\ud p~  p^{2}\phi_{esc}(p)}{\int_{0}^{\infty}\ud p ~ p^{2}f^{0}(x_{0},p)}\,,
\end{equation}
while the anisotropy of the escaping particles reads
\begin{equation}
g_{esc}(\mu)=\frac{2\vartheta(\mu)}{2+A}\left(1+A\mu \right)\,.
\end{equation}
where $\vartheta(\mu)$ is the usual Heaviside step function.

This procedure allows one to work out the expected anisotropy close to the escape boundary within the numerical approach of KJ, where the boundary condition $f(x>x_{0},p)=0$ is imposed and thus  $f(x_{0},p)$ is finite, but also in the semi-analytical approach of CAB, where $f(x_{0},p)$ is imposed to be exactly 0, simply by taking the left limit for $x\to x_{0}$ in the equations above.   

The results are shown in the right panel of Fig.~\ref{fig:escape}, along with the result of the Monte Carlo simulation by EV at $x_{0}$ (top panel) and at $x=0.99x_{0}$ (bottom panel). 
Also in this case the agreement among the different approaches is remarkable and, most interestingly, the degree of anisotropy which the methods based on the solution of the diffusion-convection equation are able to retain is indeed very accurate, even when compared to the Monte Carlo method which deals with the fully anisotropic distribution function and which can track the smooth transition from the linear to the step-like functional form of $g(x,\mu)$ in the range $0.99<x/x_{0}<1$.

Quantitatively speaking, we can safely conclude that, for plane non-relativistic shocks, the discrepancies in the spectra of accelerated and escaping particles between time-dependent and stationary approaches or between NLDSA models taking or not into account the anisotropy of the distribution function are well within any uncertainty induced, for instance, by any observational constraint of the several environmental parameters of a given astrophysical object.

\subsection{Hydrodynamics}
\begin{figure}
   \centering
   \includegraphics[width=0.48\textwidth]{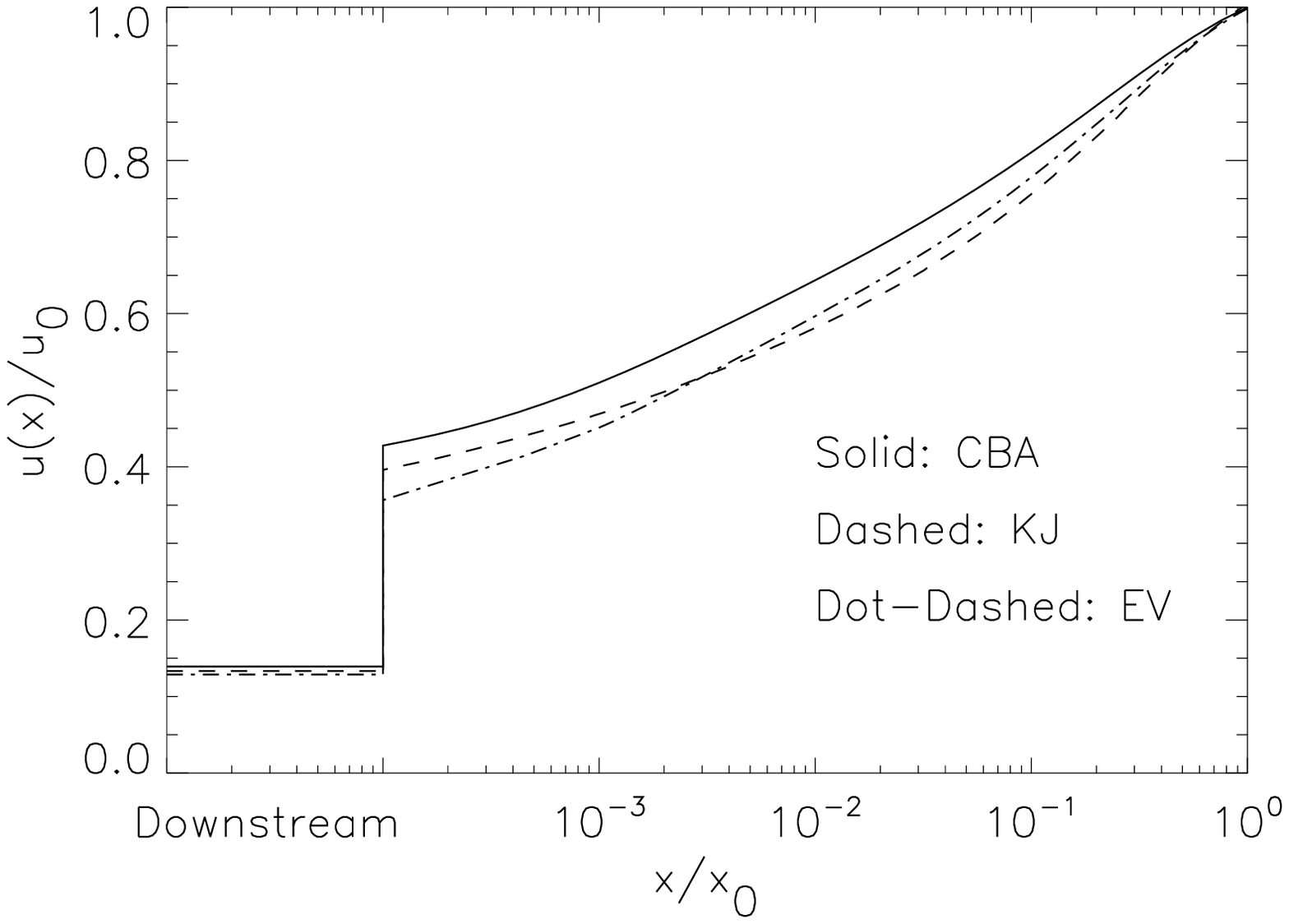}
   \includegraphics[width=0.48\textwidth]{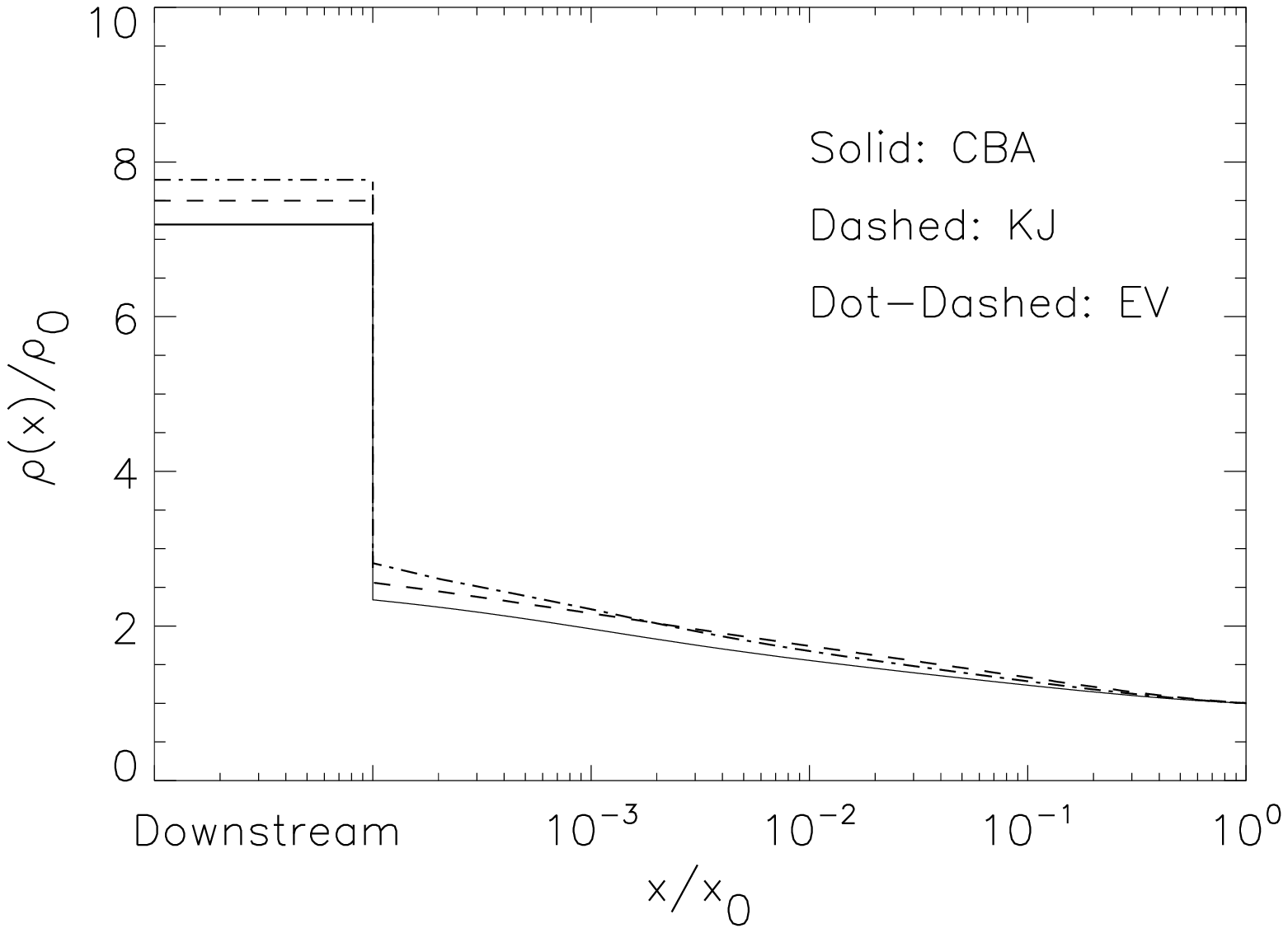}
   \caption{Spatial dependence of the fluid velocity (normalized to $u_0$, \textit{left panel}) and density (normalized to $\rho_0$\textit{right panel}) in the three different approaches.
    Here and in the following figures the shock faces to the right and is rest at $x=0$; the upstream is truncated at $x=10^{-4}x_{0}$ and the downstream is assumed as homogeneous and not in scale.}
   \label{fig:hydro}
\end{figure}

The spatial structures of the modified shocks are quite consistent among the three models. 
This is evident from Fig.~\ref{fig:hydro} which compares velocity and density profiles.
As stated above, the CR acceleration efficiencies are very close to the upper limits allowed by the system when turbulent heating is taken into account. The slightly different saturation levels achieved in different approaches might thus be regarded as due to intrinsic properties of the models.

All the models return the same ratio of unperturbed to downstream fluid density 
$\Rt=\rho_{2}/\rho_{0}$ within a tolerance of less than 10 per cent.
In particular, the total compression ratios are found to be $R_{tot,\rm{CAB}}\simeq7.2$, $R_{tot,\rm{KJ}}\simeq7.3$ and $R_{tot,\rm{EV}}\simeq7.6$, corresponding to CR acceleration efficiencies of about 60 per cent of the bulk pressure for CAB and KJ approaches and of about 75 per cent in the EV case.
The higher shock modification in the EV case is due to a larger amount of energy in 
supra-thermal, non relativistic particles compared to the CAB and KJ approaches
(see Fig.~\ref{fig:spectra}). 

In the precursor the total spread among the model profiles is within a 10-20 per cent tolerance. 
Unfortunately, at the moment there are no observations able to resolve the spatial structure of a precursor.
In addition, all the hydrodynamical evidence of efficient CR acceleration at SNR shocks comes from estimates of the separation distance between the the forward shock and the contact discontinuity \citep[see e.g.][for Tycho and SN1006, respectively]{warren,gamil}, or by an estimated suppression of the downstream plasma temperature \citep[e.g.][]{deb00} with respect to the pure gaseous case.
On the other hand, it was suggested that the H$\alpha$ profiles in Tycho's SNR shock
might indicate the presence of a CR precursor \citep[e.g.][]{w+09}.

\begin{figure}
   \centering
   \includegraphics[width=0.48\textwidth]{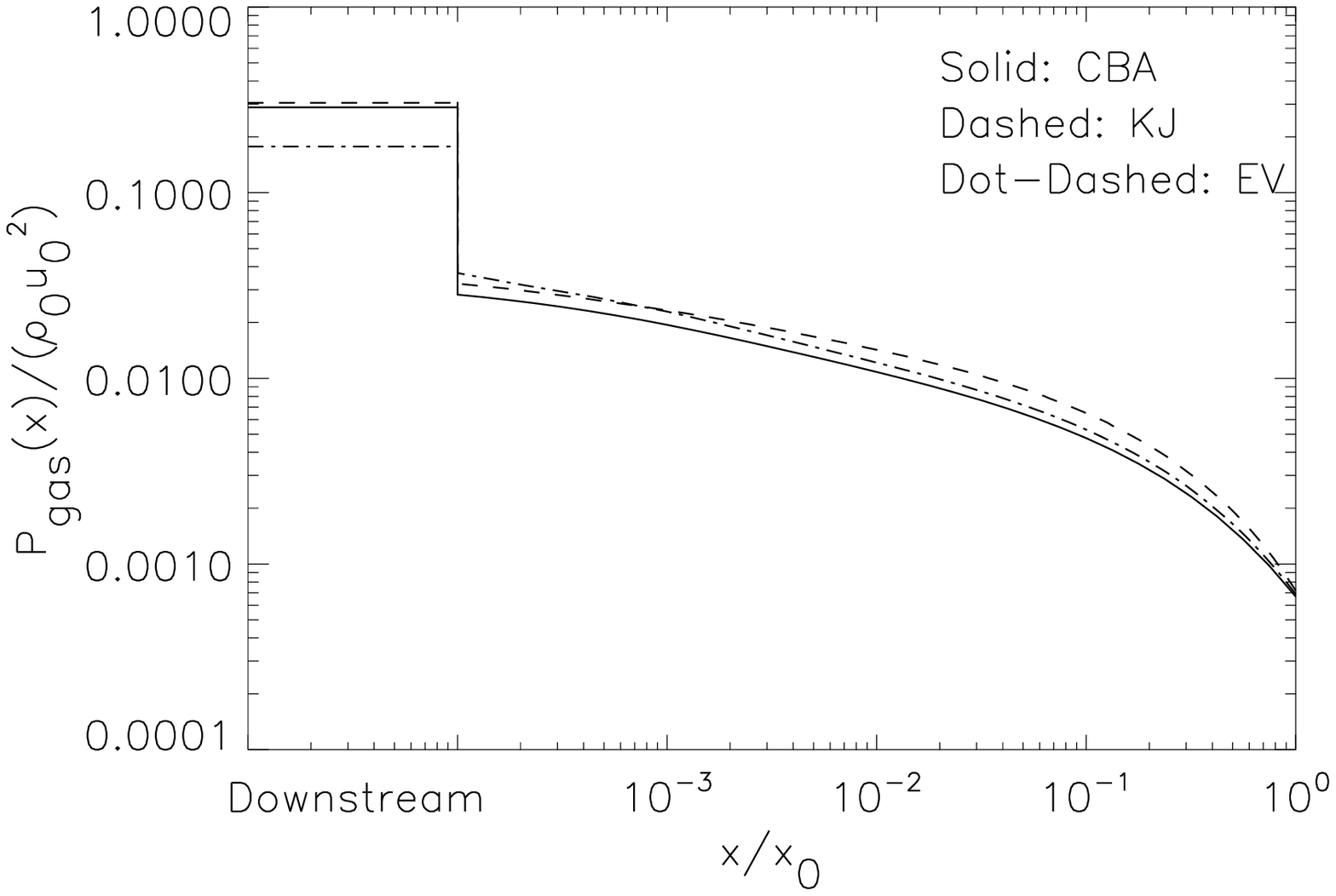}
   \includegraphics[width=0.48\textwidth]{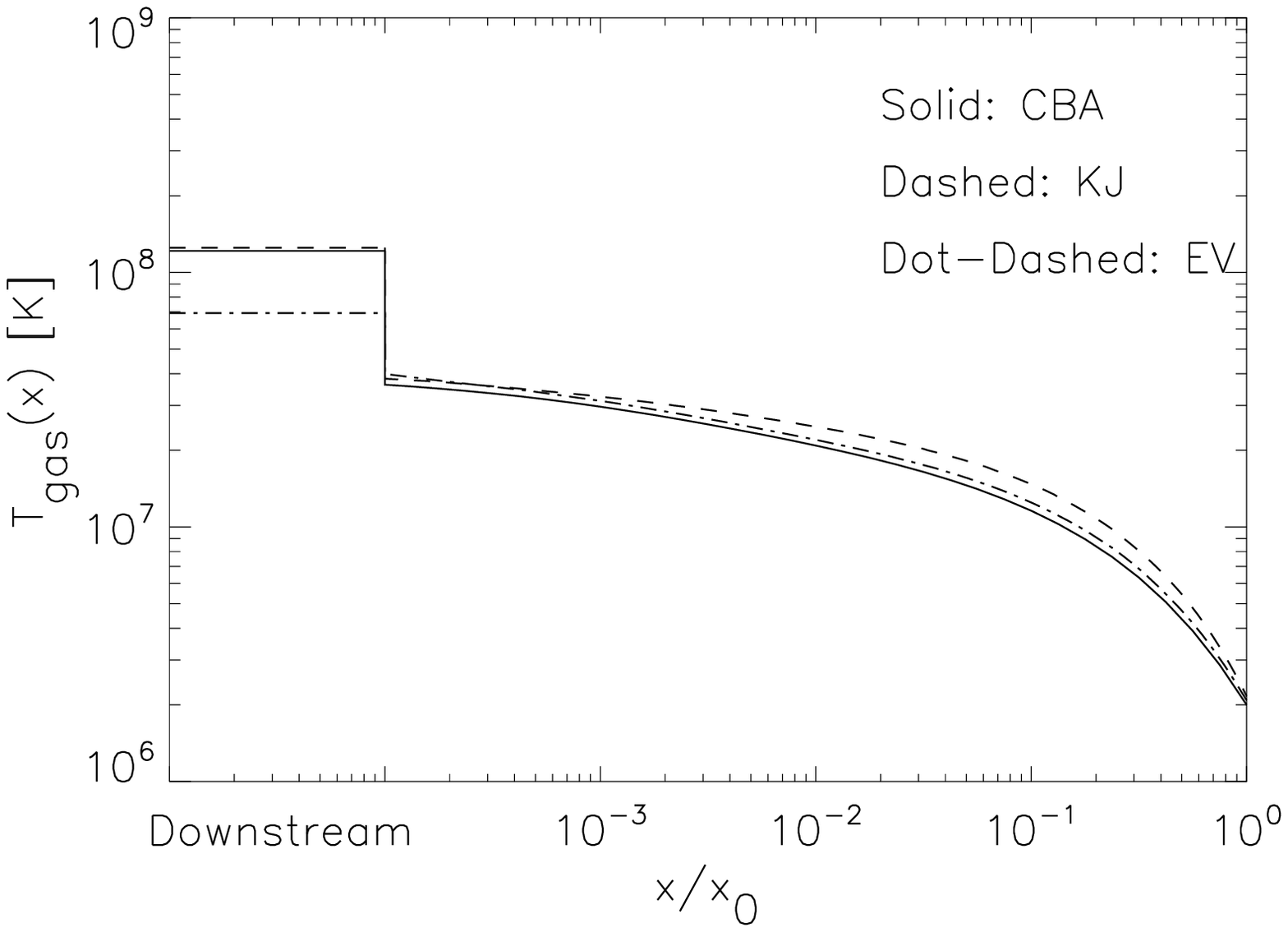}
   \caption{Spatial dependence of the gas pressure (\textit{left panel}) and temperature (\textit{right panel}) in the three different approaches. The spatial coordinate is taken as in Fig.~\ref{fig:hydro}}
   \label{fig:termo}
\end{figure}

In Fig.~\ref{fig:termo} the gas pressure and temperature are shown for the three models, confirming once again that all the approaches lead to consistent results.
The lower values of the downstream gas pressure and temperature in the Monte Carlo approach with respect to the other cases may be due to the slightly more marked shock modification, since the larger is the amount of energy channelled into accelerated particles, the lower is the energy left for heating up the downstream plasma.
On the other hand, above the thermal peak the Monte Carlo solution shows significant deviation from a pure Maxwellian distribution, because the injection model adopted in the simulation and the nature of the calculation allow mildly accelerated particles (those having crossed the shock only a few times) to be present in the shock vicinity.

Finally, the fraction of the bulk energy density flux $\rho_{0}u_{0}^{3}/2$ carried away by particles escaping the system at the upstream boundary is found to be $F_{esc,{\rm KJ}}=0.29$, $F_{esc,{\rm EV}}=0.27$ and $F_{esc,{\rm CAB}}=0.23$, consistently with the small differences in the acceleration efficiencies outlined above.

\section{Comments and conclusions}\label{sec:conclusion}
In this work we conducted a comparison among numerical, Monte Carlo and semi-analytical approaches to NLDSA theory by solving a benchmark case for an astrophysical plane, non-relativistic strong shock. 
In particular we examined a case in which cosmic rays are efficiently accelerated and strongly modify the shock with their dynamical back-reaction.

The results obtained with the three different techniques that assume specific and distinct recipes for physical ingredients such as particle injection, particle escape and diffusion properties are very consistent with each other in terms of both accelerated (and escaping) spectra and hydrodynamics.  
These findings represent an actual investigation of the role that specific pieces of information embedded in each approach have in modeling cosmic ray modified shocks.
In particular, we showed how the solution obtained when the stationary state is achieved by (a) letting the system evolve at the presence of a free-escape boundary (the numerical approach, see\S\ref{sec:KJ}), (b) a steady-state solution in which the particle transport is described statistically through a Monte Carlo technique (\S\ref{sec:EV}), or (c) the semi-analytic solution of the stationary diffusion-convection equation for the isotropic part of the cosmic ray distribution function (\S\ref{sec:CAB}) show an agreement well within any uncertainty intrinsic to the basic physics of the process. 
Thus, the agreement would not allow distinction of the methods using current observations.

At the same time, it is worth recalling that all the models are rather flexible and may be updated with any new physical insight which may come from present and future observations.
Very generally, these methods are limited more by our real knowledge of the mechanisms going on in a SNR shock than by the possibility of implementation of these effects.

An important example of a piece of information that can be embedded in all the methods is a more detailed description of the magnetic field responsible for the particle scattering.
In the calculations above we did not include any process leading to the amplification of  the background magnetic field due to CR streaming instabilities \citep{bell78, bell04}, nor any dynamical feedback of the magnetic turbulence in the shock dynamics \citep{cbavlett}.
On the other hand, by assuming Bohm-like diffusion we have intrinsically taken into account a magnetic configuration quite different from the typical interstellar one \citep{lc83a,lc83b}. 
Moreover, we included a phenomenological model for a non-adiabatic heating of the upstream background plasma as due to the instantaneous damping of all the magnetic turbulence produced via resonant streaming instability, as described in \cite{mck-v82}.  
This very efficient damping, while largely adopted in order to prevent dramatic shock modification, is however inconsistent with magnetic fields of order $100-500\mu$G inferred in the downstream of some young SNRs \citep[see e.g.][]{P+06}: in these environments the shock modification may be controlled by the non-negligible dynamical role of the amplified magnetic field, an effect which can be easily included in the models above \citep[see][]{veb06,cabv09}.

Another interesting finding of our comparison is that not only Monte Carlo simulations, but also methods based on the solution of the diffusion-convection equation can retain reliable information about the anisotropy of the distribution function, even close to the free-escape boundary, in many astrophysical cases.
This fact allows one to use the so-derived diffusive CR current to calculate the excitation of modes relevant for the growth of the magnetic turbulence, as predicted in many scenarios involving the self-generation of an amplified magnetic field responsible in which particles are scattered very efficiently \citep[see e.g.][for the calculation of the Bell non-resonant streaming instability excited by escaping particles]{reville}.

In this work we considered a situation in which injection occurs very efficiently, as suggested by the Monte Carlo modeling of the subshock.
This choice, while on one hand perfectly reproduces the observations of the Earth bow shock \citep[e.g.][]{emp90}, on the other hand is not firmly supported by observational evidence in SNR shocks.
For typical values of temperatures and magnetic fields in SNR environments, in fact, the region of the spectrum between the thermal and the non-thermal distributions does not provide, at the moment, any strong signature testable with multi-wavelength studies of SNRs.
In this respect, a simple parametrization of the injection process like the thermal leakage captures most of the (non-linear) physics which is expected to be relevant in NLDSA theory.
Further improvements in discrimination between the various models of particle injection can be expected from the ongoing work on particle-in-cell simulations of non-relativistic shocks.

Dealing with strongly modified shocks is indeed very challenging for fully numerical methods.
When huge gradients are found in the precursor, the system very likely develops some instabilities, whose nature may be not only numerical, but also physical.  For instance, the acoustic instability studied by \cite{df86,kjr92,w+09} can develop strong inhomogeneities, which can drive turbulence in multi-dimensional flows. 
Numerical methods are thus tools necessary in order to unravel these kind of questions, being in principle able to follow the growth of instabilities even in the non-linear regime, typically inaccessible to analytical approaches.

A numerical simulation is currently the only way to deal with time-dependent NLDSA problems.
We recall that there are two main ways of accounting for the dependence on time of the physical quantities: on one hand we have cases in which the shock velocity is almost constant and only CR-related quantities are expected to vary on time scales comparable with the acceleration time at $p_{max}$, and on the other hand we have cases in which also ``environmental'' quantities, like the shock velocity, change with time.

The former case is realized for instance during the ejecta dominated stage of a SNR, in which the shock is not dramatically slowed down by the inertia of the swept-up material.
Nevertheless, if a spatial boundary condition is imposed in such a way that particles are free to escape from an upstream boundary, also in this case a steady-state solution consistent with stationary approaches may be recovered. 
Otherwise, if $p_{max}$ is free to grow unconstrained by spatial limits, the system evolves according to a self-similar behavior as shown by \cite{kj07,krj09}: also in this case stationary approaches, with a proper age-limited maximum momentum \citep[see]{bac07}, should be able to reproduce the correct results. 

The latter case, instead, is expected to be realized in the Sedov-Taylor stage of the evolution of SNRs.
As shown by \cite{escape}, during this stage the decrease of the shock velocity is expected to lead to a decrease of the magnetic field amplification and in turn to a drop of the confinement power of the system. At any given time particles with momentum close to $p_{max}(t)$ cannot diffuse back to the shock and escape the system upstream, thus, carrying away a sizable fraction of the energy flow into the shock.
This situation is thought to lead to the achievement of a quasi-stationary configuration in which particle acceleration and losses balance themselves; also in this case a time-independent approach might be a good description of the problem.

In the case in which the assumption of quasi-stationarity is satisfied, modeling an astrophysical object might require one to run the NLDSA code many times: one for each time-step, for any given choice of the many environmental and model parameters.
It is clear that such an investigation can be conducted only by using very fast solutions of the NLDSA problem, and semi-analytical approaches can do this job very well \citep[see e.g.][for an example]{cab10}.

Finally, Monte Carlo codes allow one to take into account strong anisotropies of the CR distribution function, for which the diffusive approximation and hence the solution of the diffusion-convection equation for the isotropic part of $F(x,{\bf p})$ only may fail by a large amount.
This point is of central interest especially for the study of mildly-relativistic and relativistic shocks \citep{ed02}, since anisotropic corrections to Eq.~\ref{diffcon} are of order $\mathcal{O}(u^{2}/c^{2})$, as outlined in \S\ref{sec:escape}.

The last consideration is about an assumption that is ubiquitous in NLDSA theories; namely, that particle are efficiently scattered by magnetic irregularities and in a \textit{diffusive} way.
This assumption may be broken close to $x_{0}$, beyond which particles are expected to stream away without diffusing, or even in the precursor if the topology of the magnetic field is organized in structures like the cavities predicted by Bell's instability \citep{bell04,bell05}.  
Nonetheless it is worth stressing that in the literature no non-linear particle acceleration methods able to relax the hypothesis that particle are scattered in a diffusive way (which is embedded also in Monte Carlo approaches) have been developed yet.
Such a problem could be in principle addressed by Particle in Cell (PIC) simulations, but the computational time required to run a case in which the spectrum of accelerated particles spans many order of magnitudes is now, and it is going to be for years, prohibitive even for huge computer clusters.

\section*{Acknowledgments}
This research was supported in part by the National Science Foundation under Grant No. PHY05-51164, in the context of the program \textit{Particle Acceleration in Astrophysical Plasmas}, (2009, July 26-October 3), held at the Kavli Institute for Theoretical Physics (Santa Barbara, CA).
The authors want to thank all the participants to the program, but in particular way Don Ellison and Pasquale Blasi for their suggestions and illuminating contributions to this comparison project.

DC was supported by ASI through contract ASI-INAF I/088/06/0.
HK was supported by National Research Foundation of Korea through grant 2010-016425
AV was supported by NASA grants 06-ATP06-21 and NNX09AC15G.
TWJ was supported by NASA grant NNX09AH78G and by the University of Minnesota Supercomputing Institute.

\label{lastpage}

\end{document}